\documentclass[aps,prd,groupedaddress,preprint,tightenlines,eqsecnum,nofootinbib,
showpacs]{revtex4}
\usepackage{graphicx,epsf,amssymb,amsbsy,amsfonts,amssymb,amsmath,mathrsfs}
\usepackage[usenames, dvipsnames]{color}
\newif\ifdraft
\drafttrue
\newif\ifpreprint
\preprinttrue

\def\sect#1{section~{\ref{#1}}}
\def\fig#1{fig.~{\ref{#1}}}

\def\app#1{appendix~{\ref{#1}}}
\def\eqn#1{eq.~(\ref{#1})}

\def\eqns#1#2{eqs.~(\ref{#1}) and~(\ref{#2})}

\def\tab#1{table~{\ref{#1}}}

\def\P{{\rm P}}
\def\NP{{\rm NP}}

\def\L{\left(}\def\R{\right)}
\def\LB{\left[}\def\RB{\right]}
\def\Li{\mathop{\rm Li}\nolimits}
\def\Ls{\mathop{\rm Ls}\nolimits}
\def\Ll{\mathop{\rm L}\nolimits}
\def\Atree{A^{\rm tree}}
\def\matter{{\rm mat.}}

\def\spa#1.#2{\left\langle#1\,#2\right\rangle}
\def\spb#1.#2{\left[#1\,#2\right]}
\def\spash#1.#2{\spa{\smash{#1}}.{\smash{#2}}}
\def\spbsh#1.#2{\spb{\smash{#1}}.{\smash{#2}}}
\def\sand#1.#2.#3{%
\left\langle\smash{#1}{\vphantom1}^{-}\right|{#2}%
\left|\smash{#3}{\vphantom1}^{-}\right\rangle}
\def\sandpp#1.#2.#3{%
\left\langle\smash{#1}{\vphantom1}^{+}\right|{#2}%
\left|\smash{#3}{\vphantom1}^{+}\right\rangle}
\def\sandpm#1.#2.#3{%
\left\langle\smash{#1}{\vphantom1}^{+}\right|{#2}%
\left|\smash{#3}{\vphantom1}^{-}\right\rangle}
\def\sandmp#1.#2.#3{%
\left\langle\smash{#1}{\vphantom1}^{-}\right|{#2}%
\left|\smash{#3}{\vphantom1}^{+}\right\rangle}

\def\tree{{\rm tree}}
\def\oneloop{{\rm 1\hbox{-}loop}} 
\def\twoloop{{\rm 2\hbox{-}loop}} 
\def\Loop{{\rm loop}}

\def\Tr{\, {\rm Tr}}
\def\tr{\, {\rm tr}}

\def\eps{\epsilon}
\def\e{\epsilon}

\def\nn{\nonumber}

\def\n{{\tilde n}}
\def\f{\tilde f}

\def\NeqSix{{{\cal N}=6}}
\def\NeqFive{{{\cal N}=5}}
\def\NeqFour{{{\cal N}=4}}
\def\NeqTwo{{{\cal N}=2}}
\def\NeqZero{{{\cal N}=0}}

\def\NeqEight{{{\cal N}=8}}
\def\NeqSeven{{{\cal N}=7}}
\def\NeqOne{{{\cal N}=1}}

\def\scalar{{\rm scalar}}

\def\be{\begin{equation}}
\def\ee{\end{equation}}
\def\bea{\begin{eqnarray}}
\def\eea{\end{eqnarray}}
\def\ba{\begin{eqnarray}}
\def\ea{\end{eqnarray}}

\def\Ord{{\cal O}}

\DeclareMathAlphabet{\mathpzc}{OT1}{pzc}{m}{it}

\newbox\charbox
\newbox\slabox
\def\s#1{{      
        \setbox\charbox=\hbox{$#1$}
        \setbox\slabox=\hbox{$/$}
        \dimen\charbox=\ht\slabox
        \advance\dimen\charbox by -\dp\slabox
        \advance\dimen\charbox by -\ht\charbox
        \advance\dimen\charbox by \dp\charbox
        \divide\dimen\charbox by 2
        \raise-\dimen\charbox\hbox to \wd\charbox{\hss/\hss}
        \llap{$#1$} }}

\begin{document}
\hfuzz 20pt

\ifpreprint
UCLA/11/TEP/108 $\null\hskip0.1cm\null$ \hfill 
Saclay--IPhT--T11/153 $\null\hskip0.1cm\null$ \hfill
SLAC-PUB-14495                              $\null\hskip0.1cm\null$ \hfill
NSF-KITP-11-118\\
\fi

\title{$\mathcal{N} \ge 4$ Supergravity Amplitudes from Gauge Theory
at One Loop}

\author{Z.~Bern${}^a$, C. Boucher-Veronneau${}^b$, and
 H.~Johansson${}^c$}

\affiliation{
${}^a$Department of Physics and Astronomy, UCLA, Los Angeles, CA 90095, USA\\
${}^b$SLAC National Accelerator Laboratory,
              Stanford University,
             Stanford, CA 94309, USA \\
${}^c$Institut de Physique Th\'eorique, CEA--Saclay,
          F--91191 Gif-sur-Yvette cedex, France\\
$\null$\\
$\null$\\
}


\begin{abstract}
We expose simple and practical relations between the integrated four-
and five-point one-loop amplitudes of ${\cal N} \ge 4$ supergravity
and the corresponding (super-)Yang-Mills amplitudes. The link between the
amplitudes is simply understood using the recently uncovered duality
between color and kinematics that leads to a double-copy structure for
gravity. These examples provide additional direct confirmations of
the duality and double-copy properties at loop level for a sample of
different theories.
\end{abstract}

\pacs{04.65.+e, 11.15.Bt, 11.30.Pb, 11.55.Bq \hspace{1cm}}

\maketitle


\section{Introduction}

One of the remarkable theoretical ideas emerging in the last decade is
the notion that gravity theories are intimately tied with gauge
theories. The most celebrated connection is the AdS/CFT
correspondence~\cite{AdSCFT} which relates maximally
supersymmetric Yang-Mills gauge theory to string theory (and
supergravity) in anti-de Sitter space.  Another surprising link
between the two theories is the conjecture that to all perturbative
loop orders the kinematic numerators of diagrams describing gravity
scattering amplitudes are double copies of the gauge-theory
ones~\cite{BCJ,BCJLoop}.  This double-copy relation relies on a novel
conjectured duality between color and kinematic diagrammatic
numerators of gauge-theory scattering amplitudes. At tree level, the
double-copy relation encodes the Kawai-Lewellen-Tye (KLT) relations
between gravity and gauge-theory amplitudes~\cite{KLT}.

The duality between color and kinematics offers a powerful tool for
constructing both gauge and gravity loop-level scattering amplitudes,
including nonplanar contributions~\cite{BCJLoop, JJHenrikReview, ck4l,
  fivepointBCJ}.  The double-copy property does not rely on
supersymmetry and is conjectured to hold just as well in a wide
variety of supersymmetric and non-supersymmetric theories.  In recent
years there has been enormous progress in constructing planar
$\NeqFour$ super-Yang-Mills amplitudes.  For example, at four and five
points, expressions for amplitudes of this theory---believed to be
valid to all loop orders and nonperturbatively---have been
constructed~\cite{BDS}.  (For recent reviews, see
refs.~\cite{AmplitudeReviews,JJHenrikReview}.)  Many of the new
advances stem from identifying a new symmetry, called dual conformal
symmetry, in the planar sector of $\NeqFour$ super-Yang-Mills
theory~\cite{DualConformal}. This symmetry greatly enhances the power
of methods based on
unitarity~\cite{UnitarityMethodI,UnitarityMethodII} or on recursive
constructions of integrands~\cite{BCFWLoops}.  The nonplanar sector of
the theory, however, does not appear to possess an analogous symmetry.
Nevertheless, the duality between color and kinematics offers a
promising means for carrying advances in the planar sector of
$\NeqFour$ super-Yang-Mills theory to the nonplanar sector and then to
$\NeqEight$ supergravity.  In particular, the duality interlocks
planar and nonplanar contributions into a rigid structure.  For
example, as shown in ref.~\cite{BCJLoop}, for the three-loop
four-point amplitude, the maximal cut~\cite{MaximalCut} of a single
planar diagram is sufficient to determine the complete amplitude,
including nonplanar contributions.

Here we will explore one-loop consequences of the duality between
color and kinematics for supergravity theories with $4 \le {\cal N}\le
6$ supersymmetries.  These cases are less well understood than the
cases of maximal supersymmetry. (Some consequences for finite one-loop
amplitudes in non-supersymmetric pure Yang-Mills theory have been
studied recently~\cite{FiniteKKandBCJRelations}.)  Since the duality
and its double-copy consequence remain a conjecture, it is an
interesting question to see if the properties hold in the simplest
nontrivial loop examples with less than maximal supersymmetry.  In
particular, we will explicitly study the four- and five-point
amplitudes of these theories.  These cases are especially
straightforward to investigate because the required gauge theory and
gravity amplitudes are known. Our task is then to find rearrangements
that expose the desired properties. The necessary gauge-theory
four-point amplitudes were first given in dimensional regularization
near four dimensions in ref.~\cite{FDH}, and later in a form valid to
all orders in the dimensional regularization
parameter~\cite{BernMorgan}.  At five points, the dimensionally
regularized gauge-theory amplitudes near four dimensions were
presented in ref.~\cite{FiveYM}.  The four-graviton amplitudes in
theories with ${\cal N} \le 6 $ supersymmetries were first given in
ref.~\cite{DunbarNorridge}.  More recently, the MHV one-loop
amplitudes of $\NeqSix$ and $\NeqFour$ supergravity were presented, up
to rational terms in the latter
theory~\cite{DunbarSugra}.\footnote{While completing the present
paper, version 2 of ref.~\cite{DunbarSugra} appeared, giving the
missing rational terms of the $\NeqFour$ supergravity five-point
amplitudes.}

Here we point out that the double-copy relations can be
straightforwardly exploited, allowing us to obtain complete integrated
four- and five-point amplitudes of ${\cal N} \ge 4$ supergravity
amplitudes as a simple linear combinations of corresponding
gauge-theory amplitudes. Because these relations are valid in any
number of dimensions, we can use previously obtained representations
of QCD and $\NeqFour$ super-Yang-Mills four-point amplitudes valid
with $D$-dimensional momenta and states in the loop to obtain such
representations for ${\cal N} \ge 4$ supergravity. These
$D$-dimensional results are new, while our four-dimensional results
reproduce ones found in refs.~\cite{DunbarNorridge,DunbarSugra}.
Relations between integrated $\NeqFour$ super-Yang-Mills and
$\NeqEight$ supergravity four-point one- and two-loop amplitudes had
been described previously in refs.~\cite{SchnitzerSubleadingColor}.

For cases with larger numbers of external legs, the loop momentum is
expected to become entangled with the relations making them more
intricate.  Nevertheless, we expect that the duality should lead to 
simple structures at one loop for all multiplicity, and once
understood these should lead to improved means for constructing
gravity loop amplitudes.  Indeed, the duality has already been enormously
helpful for constructing four- and five-point multiloop
amplitudes in $\NeqEight$
supergravity~\cite{BCJLoop,JJHenrikReview,fivepointBCJ, ck4l}.

This paper is organized as follows.  In \sect{ReviewSection} we review
some properties of scattering amplitudes, including the conjectured
duality between color and kinematics and the gravity double-copy property.
Then in \sect{OneLoopImplicationsSection}, we give some one-loop
implications, before turning to supergravity.  We also make a few
comments in this section on two-loop four-point amplitudes. We give
our summary and outlook in \sect{ConclusionSection}.  Two appendices
are included collecting gauge-theory amplitudes and explicit forms of the
integrals used in our construction.

\section{Review}
\label{ReviewSection}

In this section we review some properties of gauge and gravity amplitudes
pertinent to our construction of supergravity amplitudes.  We first
summarize the duality between color and kinematics which allows us
to express gravity amplitudes in terms of gauge-theory ones.  We then
review decompositions of one-loop ${\cal N} = 4,5,6$ supergravity
amplitudes in terms of contributions of matter multiplets, simplifying
the construction of the amplitudes.

\subsection{Duality between color and kinematics}

We can write any $m$-point $L$-loop-level gauge-theory amplitude where all
particles are in the adjoint representation as
\begin{equation}
  \frac{(-i)^L}{g^{m-2 +2L }}{\cal A}^\Loop_m = 
 \sum_{j}{\int \frac{d^{DL} p}{ (2 \pi)^{DL}}
  \frac{1}{S_j}  \frac {n_j c_j}{\prod_{\alpha_j}{p^2_{\alpha_j}}}}\,.
\label{LoopGauge} 
\end{equation}
The sum runs over
the set of distinct $m$-point $L$-loop graphs, labeled by $j$, with
only cubic vertices, corresponding to the diagrams of a $\phi^3$
theory.  The product in the denominator runs over all Feynman
propagators of each cubic diagram.  The integrals are over $L$
independent $D$-dimensional loop momenta, with measure $d^{DL}
p=\prod_{l=1}^{L}d^Dp_l$.  The $c_i$ are the color factors obtained by
dressing every three vertex with an $\f^{abc} = i \sqrt{2}
f^{abc}=\Tr\{[T^{a},T^{b}]T^{c}\}$ structure constant, and the $n_i$
are kinematic numerator factors depending on momenta, polarizations
and spinors.  For supersymmetric amplitudes expressed in superspace,
there will also be Grassmann parameters in the numerators.  The $S_j$
are the internal symmetry factors of each diagram.  The form in
\eqn{LoopGauge} can be obtained in various ways; for example, starting
from covariant Feynman diagrams, where the contact terms are absorbed
into kinematic numerators using inverse propagators.

Any gauge-theory amplitude of the form~(\ref{LoopGauge}) possesses an invariance under ``generalized gauge
transformations''~\cite{BCJ,Tye,BCJLoop,Square,VamanYao} corresponding
to all possible shifts, $n_i \rightarrow n_i + \Delta_i$, where the
$\Delta_i$ are arbitrary kinematic functions (independent of color)
constrained to satisfy
\begin{equation}
\sum_{j} \int  \frac{d^{DL} p}{(2 \pi)^{DL}}
\frac{1}{S_j} \frac{\Delta_j c_j}{\prod_{\alpha_j}{p^2_{\alpha_j}}} = 0\,.
\label{gaugeInvar}
\end{equation}
By construction this constraint ensures that the shifts by $\Delta_i$ do not
alter the amplitude (\ref{LoopGauge}).  The condition (\ref{gaugeInvar}) can be satisfied either because of algebraic identities of the integrand (including identities obtained after trivial relabeling of loop momenta in diagrams) or because of nontrivial integration identities.
Here we are interested in $\Delta_i$ that satisfy (\ref{gaugeInvar}) because of the former reason, as the relations we will discuss below operate at the integrand level. We will refer to these kind of numerator shifts valid at the integrand level as point-by-point generalized gauge transformations.
One way to express this freedom is by taking any function of the
momenta and polarizations and multiplying by a sum of color factors
that vanish by the color-group Jacobi identity, and then repackaging the functions into $\Delta_i$'s  over propagators according to the color factor of each individual term.  Some of the resulting freedom corresponds to gauge
transformations in the traditional sense, while most does not.  These
generalized gauge transformations will play a key role, allowing us to 
choose different representations of gauge-theory amplitudes, aiding
our construction of gravity amplitudes from gauge-theory ones.

The conjectured duality of refs.~\cite{BCJ,BCJLoop} states that to all
loop orders there exists a form of the amplitude where triplets of
numerators satisfy equations in one-to-one correspondence with the
Jacobi identities of the color factors,
\begin{equation}
c_i = c_j - c_k \;  \Rightarrow \;  n_i = n_j - n_k \,,
\label{BCJDuality}
\end{equation}
where the indices $i,j,k$ schematically indicate the diagram to which
the color factors and numerators belong to.  Moreover, we demand that
the numerator factors have the same antisymmetry property as color
factors under interchange of two legs attaching to a cubic vertex,
\begin{equation}
c_i \rightarrow - c_i \;  \Rightarrow \;  n_i \rightarrow - n_i \,.
\label{BCJAntiSymmetry}
\end{equation}

At tree level, explicit forms satisfying the duality have been given
for an arbitrary number of external legs and any helicity
configuration~\cite{TreeAllN}.  An interesting consequence of this
duality is nontrivial relations between the color-ordered partial
tree amplitudes of gauge theory~\cite{BCJ} which
have been proven in gauge theory~\cite{Feng} and in string
theory~\cite{Bjerrum1}. Recently these relations played an important
role in the impressive construction of the complete solution to all
open string tree-level amplitudes~\cite{AllStringAmplitudes}.  The
duality has also been studied from the vantage point of the heterotic
string, which offers a parallel treatment of color and
kinematics~\cite{Tye}.  A partial Lagrangian understanding of the
duality has also been given~\cite{Square}.  The duality
(\ref{BCJDuality}) has also been expressed in terms of an alternative
trace-based representation~\cite{Trace}, emphasizing the underlying
group-theoretic structure of the duality.  Indeed, at least for
self-dual field configurations and MHV amplitudes, the underlying
infinite-dimensional Lie algebra has been very recently been
identified as area preserving diffeomorphisms~\cite{OConnell}.

At loop level, less is known though some nontrivial tests have been
performed. In particular, the duality has been confirmed to hold for
the one-, two- and three-loop four-point amplitudes of $\NeqFour$
super-Yang-Mills theory~\cite{BCJLoop}.  It is also known to hold for
the one- and two-loop four-point identical helicity amplitudes of pure
Yang-Mills theory~\cite{BCJLoop}. Very recently it has also been shown to 
hold for the four-loop four-point amplitude of $\NeqFour$
super-Yang-Mills theory~\cite{ck4l}, and for the five-point one-, two-
and three-loop amplitudes of the same theory~\cite{fivepointBCJ}. 

\subsection{Gravity as a double copy of gauge theory}

Perhaps more surprising than the gauge-theory aspects of the duality between color
and kinematics is a directly related conjecture for the detailed structure of gravity amplitudes. Once the gauge-theory
amplitudes are arranged into a form satisfying the duality
(\ref{BCJDuality}), corresponding gravity amplitudes can be obtained
simply by taking a double copy of gauge-theory numerator
factors~\cite{BCJ,BCJLoop},
\begin{equation}
 {\frac{(-i)^{L+1}}{(\kappa/2)^{n-2+2L}}} \! {\cal M}^\Loop_m = 
\sum_{j} {\int \frac{d^{DL} p}{(2 \pi)^{DL}}
 \frac{1}{S_j} \frac{n_j \n_j}{\prod_{\alpha_j}{p^2_{\alpha_j}}}} \, , 
\hskip .7 cm 
\label{DoubleCopy}
\end{equation}
where ${\cal M}^\Loop_m$ are $m$-point $L$-loop gravity amplitudes.  The $\n_i$
represent numerator factors of a second gauge-theory amplitude, the
sum runs over the same set of diagrams as in \eqn{LoopGauge}.  At
least one family of numerators ($n_j$ or $\n_j$) for gravity must be
constrained to satisfy the
duality~(\ref{BCJDuality})~\cite{BCJLoop,Square}.  This is expected to 
hold in a large class of gravity theories, including all theories that
are low-energy limits of string theories.  We obtain different
gravity theories by taking the $n_i$ and $\n_i$ to be numerators of
amplitudes from different gauge theories.  Here we are interested in ${\cal N} \ge 4$ supergravity amplitudes 
in $D=4$.  For example, we obtain the pure supergravity theories as products of $D=4$ Yang-Mills
theories as,
\begin{eqnarray}
&& \NeqEight\ \hbox{supergravity} : (\NeqFour\ \hbox{sYM}) \times 
                                   (\NeqFour\ \hbox{sYM})\,,\nn \\
&& \NeqSix\ \hbox{supergravity} : (\NeqFour\ \hbox{sYM}) \times 
                                   (\NeqTwo\ \hbox{sYM}) \,,\nn \\
&& \NeqFive\ \hbox{supergravity} : (\NeqFour\ \hbox{sYM}) \times 
                                   (\NeqOne\ \hbox{sYM}) \,,\nn \\
&& \NeqFour\ \hbox{supergravity} : (\NeqFour\ \hbox{sYM}) \times 
                                   (\NeqZero\ \hbox{sYM})  \,,
\end{eqnarray}
where $\NeqZero$ super-Yang-Mills is ordinary non-supersymmetric
Yang-Mills theory, consisting purely of gluons.
(${\cal N} = 7$ supergravity is equivalent to
${\cal N} = 8$ supergravity, so we do not list it.)

Since the duality requires the numerators and color factors to share
the same algebraic properties (\ref{BCJDuality}) and (\ref{BCJAntiSymmetry}), \eqn{gaugeInvar} implies that
\begin{equation}
\sum_{j}{\int \frac{d^{DL} p}{(2 \pi)^{DL}}
\frac{1}{S_j} \frac {\Delta_j \n_j}{\prod_{\alpha_j}{p^2_{\alpha_j}}}} = 0\,,
\label{GravgaugeInvar}
\end{equation}
so that the gravity amplitude (\ref{DoubleCopy}) is invariant under
the same point-by-point generalized gauge transformation $n_j \rightarrow n_j +\Delta_j$
as in gauge theory.

At tree level, the double-copy property encodes the KLT~\cite{KLT}
relations between gravity and gauge theory~\cite{BCJ}.  The
double-copy formula (\ref{DoubleCopy}) has been proven at tree level
for pure gravity and for $\NeqEight$ supergravity,
when the duality~(\ref{BCJDuality}) holds in the corresponding gauge
theories~\cite{Square}.  At loop level a simple argument based on the
unitarity cuts strongly suggests that the double-copy property should
hold if the duality holds in gauge theory~\cite{BCJLoop,Square}.  In
any case, the nontrivial part of the loop-level conjecture is the
assumption of the existence of a gauge-theory loop amplitude
representation that satisfies the duality between color and
kinematics. The double-copy property (\ref{DoubleCopy}) has been explicitly
confirmed in $\NeqEight$ supergravity through four loops for the
four-point amplitudes~\cite{BCJLoop,ck4l} and through two loops for
the five-point amplitudes~\cite{fivepointBCJ}.  (The three- and
four-loop $\NeqFour$ super-Yang-Mills and $\NeqEight$ supergravity four-point amplitudes had been
given earlier, but in a form where the duality and double copy are not
manifest~\cite{GravityThree,CompactThree,Neq44np,GravityFour}.)

\subsection{Decomposing one-loop ${\cal N} \ge 4$ supergravity amplitudes.}
\label{DeomposingSubsection}

\begin{table}[th] 
  \begin{center}
    \begin{tabular} {| c || c | c | c | c | c | }
      \hline
     & scalars & spin 1/2 & spin 1 & spin 3/2 & spin 2  \\ \hline \hline
     $\mathcal{N} = 8$ & 70 & 56 & 28 & 8 & 1 \\ \hline
      $\mathcal{N} = 6$ gravity & 30 & 26 & 16 & 6 & 1 \\ \hline
     $\mathcal{N} = 5$ gravity & 10 & 11 & 10 & 5 &1  \\ \hline
     $\mathcal{N} = 4$ gravity & 2 & 4 & 6 & 4 & 1 \\ \hline
     $\mathcal{N} = 6$ matter & 20 & 15 & 6 & 1 &  \\ \hline
     $\mathcal{N} = 4$ matter & 6 & 4 & 1 &   &\\ \hline
     
    \end{tabular}
    \caption{Particle content of relevant supergravity multiplets. 
The scalars are taken to be real for counts in this table.}
\label{MultipletsTable}
    \end{center}
  \end{table}

To simplify the analysis, we consider amplitudes with only gravitons on the
external legs. (One can, of course, use an on-shell superspace as described
in ref.~\cite{OnShellSuperSpace} to include other
cases as well.) 
At one loop it is well known that the graviton scattering amplitudes of various
supersymmetric theories satisfy simple linear relations dictated by
the counting of states in each theory.  
In \tab{MultipletsTable} we give the particle content of relevant
supergravity multiplets. (The $\NeqFive$ matter multiplet is the same as the $\NeqSix$ matter one, hence, it is not explicitly listed. Similarly, the
$\NeqEight$ supergravity multiplet is equivalent to the $\NeqSeven$ one.)
Looking at this table, we can easily assemble some simple relations
between the contributions from different multiplets circulating in the
loop,
\bea
{\cal M}^{\text{1-loop}}_{\NeqSix}(1,2,\ldots ,m)&=&{\cal M}^{\text{1-loop}}_{\NeqEight}(1,2,\ldots ,m)-2{\cal M}^{\text{1-loop}}_{\NeqSix, \matter}(1,2 ,\ldots,m)\,, \nn \\
{\cal M}^{\text{1-loop}}_{\NeqFive}(1,2, \ldots,m)&=&
{\cal M}^{\text{1-loop}}_{\NeqEight}(1,2, \ldots,m)-3{\cal M}^{\text{1-loop}}_{\NeqSix,\matter}(1,2, \ldots, m)\,, \label{SugraAssembly} \\
{\cal M}^{\text{1-loop}}_{\NeqFour}(1,2, \ldots,m)&=&{\cal M}^{\text{1-loop}}_{\NeqEight}(1,2, \ldots,m)-4{\cal M}^{\text{1-loop}}_{\NeqSix,\matter}(1,2, \ldots,m) \nn\\
&& \null \hskip 1 cm 
  + 2{\cal M}^{\text{1-loop}}_{\NeqFour, \matter}(1,2, \ldots,m)\,,\nn
 \eea
where the subscript ``mat" denotes a matter multiplet contribution.
Thus, in the rest of the paper, we will consider only
one-loop amplitudes with the two types of matter going around the
loop in addition to the $\NeqEight$ amplitudes. The remaining ${\cal N}\ge 4$ amplitudes (with generic amounts of ${\cal N}\ge 4$ matter) can be assembled by linear combination of these three types.

\section{Implications of the duality at one loop}
\label{OneLoopImplicationsSection}

In this section we first present a few general one-loop implications
of the duality between color and kinematics.  Our initial
considerations are general and apply as well to non-supersymmetric
theories. We will then specialize to ${\cal
N}\ge4$ supergravity four- and five-point amplitudes, taking advantage
of special properties of $\NeqFour$ super-Yang-Mills theory.

\subsection{Implications for generic one-loop amplitudes}

\begin{figure*}[tb] 
\begin{center}
\includegraphics[scale=0.45]{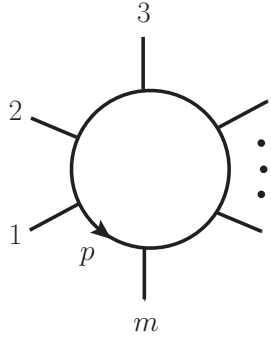}
\end{center}
\vskip -.9 cm 
\caption[a]{\small the one-loop $m$-gon master diagram for the
cyclic ordering $1,2, \ldots, m$.
}
\label{NgonParentFigure}
\end{figure*}

As shown in ref.~\cite{DixonColor} 
all color factors appearing in a one-loop amplitude can be obtained
from the color factors of ``ring diagrams'', that is the
$(m-1)!/2$ one-particle-irreducible (1PI) diagrams in the shape of a ring,
as illustrated in \fig{NgonParentFigure} for the cyclic ordering 
$1,2, \ldots, m$. We will denote the color and
kinematic numerator factors of such a diagram with external leg ordering
$1,2,\ldots, m$ by $c_{123\cdots m}$ and $n_{123\cdots m}(p)$.
Its color factor is given by the adjoint trace,
\begin{equation}
c_{123\ldots m}=\Tr_A[ \f^{a_1} \f^{a_2}\f^{a_3}\cdots \f^{a_m} ]\,,
\label{AdjointTrace}
\end{equation}
where $(\f^{a_i})^{bc}=\f^{b a_i c}$.  
%

\begin{figure*}[tb]
\begin{center}
\vskip .7 cm 
\includegraphics[scale=.43]{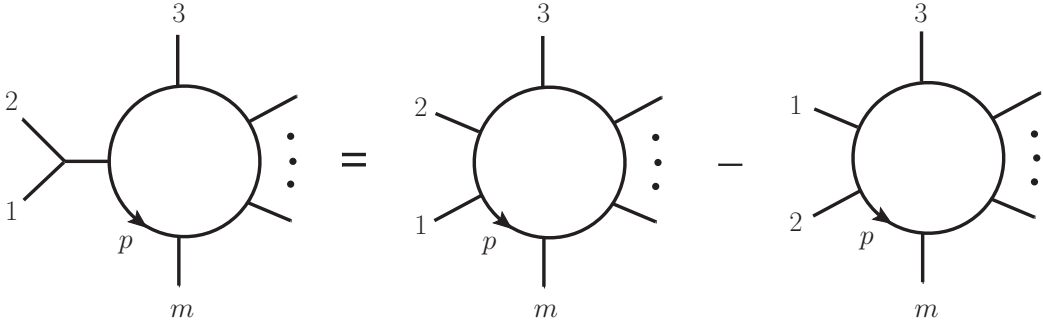}
\end{center}
\vskip -.7 cm 
\caption[a]{\small The basic Jacobi relation between three one-loop graphs
that can be used to express any color factor or kinematic numerator 
factor for any one-loop graph in terms of the parent $m$-gons. 
\label{OneloopJacobiFigure}
}
\end{figure*}

%
The color factors of the one-particle-reducible diagrams are simply
given by antisymmetrizations of ring-diagram ones as dictated by the
Jacobi relations (\ref{BCJDuality}).  For example, the color factor of
the diagram with a single vertex external to the loop shown in
\fig{OneloopJacobiFigure} is
\begin{equation}
c_{[12]3\cdots m} \equiv c_{123\cdots m}-c_{213\cdots m}\,.
\label{DJacobi}
\end{equation}
If we have a form of the amplitude where the duality holds, then the
numerator of this diagram is
\begin{equation}
n_{[12]3\cdots m}(p) \equiv n_{123\cdots m}(p)-n_{213\cdots m}(p)\,.
\label{D2Jacobi}
\end{equation}
The color factors of other diagrams, with multiple vertices
external to the loop, can similarly be obtained with further
antisymmetrizations such as $c_{[[12]3]\cdots m} = c_{[12]3\cdots
  m}-c_{3[12]\cdots m}$.  In this way all color factors and numerators
can be expressed in terms of the ones of the ring diagram, so it
serves as our ``master'' diagram.

It is also useful to consider representations where the dual Jacobi relations do not hold.  For any $m$-point one-loop amplitude, we can use the color-group Jacobi
identity to eliminate all color factors except those of the master
diagram and its relabelings. Indeed, this is how one arrives at the
adjoint-representation color basis~\cite{DixonColor}.  In this color
basis we express the one-loop amplitude in terms of a sum over
permutations of a planar integrand,
\begin{equation}
{\cal A}^\oneloop(1,2, \ldots, m) = g^m \sum_{S_m/(Z_m\times Z_2)} 
 \int \frac{d^D p}{(2 \pi)^D} \, c_{123\ldots m} \, 
\mathscr{A}(1,2, \ldots, m; p)\,,
\label{OneLoopYMGen}
\end{equation}
where $\mathscr{A}(1,2,\ldots,m;p)$ is the complete integrand of
the color-ordered amplitude, $A^\oneloop(1,2,\ldots,m)$. 
The sum runs over all permutations of external legs ($S_n$), but with the cyclic ($Z_m$) and reflection ($Z_2$) permutations modded out. In
this representation all numerator factors except for the $m$-gon ones
are effectively set to zero, since their color factors no longer appear
in the amplitude. This is equivalent to a generalized gauge
transformation applied to the numerators\footnote{Here we have
  absorbed a phase factor $i$ into the numerator definition, $i \,
  n_j\rightarrow n_j$, compared to \eqn{LoopGauge}, as is convenient 
for one-loop amplitudes. For the remaining part of the paper we will 
use this convention.}
\bea 
n_{123\cdots m}(p)&\rightarrow & n_{123\cdots
  m}(p)+\Delta_{123\cdots m}(p)=\mathscr{A}(1,2,3, \ldots, m; p)
     {\prod_{\alpha=1}^{m}{p^2_{\alpha}}}\,, \nn\\
n_i &\rightarrow &
     n_i+\Delta_i=0,~~~~~~~{\rm for~1PR~graphs}~i\,,
\label{newNumerators}
\eea
where the product $\prod p_\alpha^2$ runs over the inverse propagators
of the $m$-gon master diagram. In this representation the $m$-gon
numerators are in general nonlocal to account for propagators
carrying external momenta present in the one-particle reducible (1PR)
diagrams but not in master diagrams. In general, the new numerators
in \eqn{newNumerators} will not satisfy the duality relations
(\ref{BCJDuality}).

Recall that generalized gauge invariance implies that only one of the
two copies of numerators needs to satisfy the duality in order for the
double-copy property to work.  For the first copy we use the
duality-violating representation (\ref{newNumerators}) where all
one-particle reducible numerator factors are eliminated in favor of
nonlocal $m$-gon master numerator factors.  For the second copy we
use the duality-satisfying numerators, $\tilde{n}_{12\ldots m}$.
Then according to the double-copy formula (\ref{DoubleCopy}), by
making the substitution $c_i \rightarrow \tilde{n}_i$ in
\eqn{OneLoopYMGen}, we obtain a valid gravity amplitude.  We then have
\begin{equation}
{\cal M}^\oneloop (1,2, \ldots, m) = 
\Bigl( \frac{\kappa}{2} \Bigr)^m
\sum_{S_m/(Z_m\times Z_2)}  \int \frac{d^D p}{(2 \pi)^D} \,
\tilde{n}_{123 \dots m}(p) \, \mathscr{A}(1,2, \ldots, m; p)\,,
\label{OneloopGravityGen} 
\end{equation}
where $\tilde{n}_{12 \dots m}(p)$ is the $m$-gon master numerator with
the indicated ordering of legs and we have replaced
the gauge-theory coupling constant with the gravity one.

At first sight, it may seem surprising that only the $m$-gon numerators
are needed, but as noted above, these master numerators contain all the
nontrivial information in the amplitudes.  The nontrivial step in
this construction is to find at least one copy of $m$-gon numerators
$\tilde{n}_i$ such that the duality relations (\ref{BCJDuality}) hold
manifestly.

So far these considerations have been general. An important simplification occurs if the
numerators of one of the gauge-theory copies 
are independent of the loop momenta, $\tilde{n}_{123
  \dots m}(p)=\tilde{n}_{123 \dots m}$. We can then 
pull these numerators out of the integral in \eqn{OneloopGravityGen}
giving relations between {\it integrated} gravity and gauge theory amplitudes.
Below we will identify two
cases where this is indeed true: the four- and five- point one-loop
amplitudes of $\NeqFour$ super-Yang-Mills
theory~\cite{GSB,fivepointBCJ}.  Taking one copy to be the $\NeqFour$
super-Yang-Mills amplitude and the other to be a gauge-theory
amplitude with fewer supersymmetries, we then get a remarkably simple
relation between integrated one-loop $(\mathcal{N} + 4)$ supergravity
and super-Yang-Mills amplitudes with $\mathcal{N}$ supersymmetries,
\begin{equation}
{\cal M}_{{\cal N}+4~{\rm susy}}^\oneloop (1,2, \ldots, m) = 
\Bigl( \frac{\kappa}{2} \Bigr)^m
\sum_{S_m/(Z_m\times Z_2)}\, \tilde{n}_{123 \dots m} \, 
A_{{\cal N}~{\rm susy}}^\oneloop(1,2, \ldots, m)\,,
\label{SimpleOneloopGravitySusy} 
\end{equation}
valid for $m=4,5$.  This construction makes manifest the remarkably
good power counting noted in
refs.~\cite{UnexpectedCancellations,DunbarSugra}.  We do not expect
higher points to be quite this simple, but we do anticipate strong
constraints between generic one-loop amplitudes of gravity theories and
those of gauge theory.


\subsection{Four-point one-loop  ${\cal N} \ge 4$ supergravity amplitudes}
\label{FourPtsGravSubsection}

We now specialize the above general considerations to four-point
supergravity amplitude. There is only one independent four-graviton
amplitude, ${\cal M}_{{\cal N} \rm susy}^{\oneloop}(1^-,2^-,3^+,4^+)$, 
as the others either vanish or are trivially related by relabelings.
As a warmup exercise, we start with $\NeqEight$ supergravity and we
reevaluate this supergravity amplitude using the above considerations.
Our starting point is the $\NeqFour$ super-Yang-Mills one-loop
four-point amplitude~\cite{GSB,BRY},
\begin{equation} 
{\cal A}^{\oneloop}_{\NeqFour}(1,2,3,4)= i  s t g^4 A^\tree(1,2,3,4)
\Bigl( c_{1234} I_4^{1234} + 
       c_{1243} I_4^{1243} +
       c_{1423} I_4^{1423} \Bigr) \,,
\label{NeqFourYMFourPoint}
\end{equation}
where $s = (k_1 + k_2)^2$ and $t = (k_2 + k_3)^2$ are the usual Mandelstam 
invariants, and the tree amplitude is
\be
 A^\tree(1^-,2^-,3^+,4^+) = \frac{ i \spa{1}.{2}^4}{ \spa{1}.{2} \spa{2}.{3} 
      \spa3.4 \spa4.1}\,,
\ee
where the angle brackets $\spa{i}.{j}$ (also $\spb{i}.{j}$ below)
denotes spinor products. (See {\it e.g.} ref.~\cite{DixonTASI}.)
The function $I_4^{1234}$ is the massless scalar box integral defined in
\eqns{ZeroMassBoxFull}{ZeroMassBoxFullExpand} of
\app{IntegralsAppendix}. The other box integrals are just relabelings
of this one.  The expression in \eqn{NeqFourYMFourPoint} in terms of
the box integral (\ref{ZeroMassBoxFull}) is valid in dimensions
$D<10$.

The first color factor in \eqn{NeqFourYMFourPoint} is given by
\be
c_{1234}=\f^{ba_1c}\f^{ca_2d}\f^{da_3e}\f^{ea_4b}\,,
\ee
and the others are just relabelings of this one.
The kinematic numerator in each case is
\begin{equation}
n_{1234} = n_{1243} = n_{1423} = i st A^\tree(1,2,3,4) \,.
\label{FourFourNumeratorCrossing}
\end{equation}
These numerators happen to have full crossing symmetry, but that is a
special feature of the four-point amplitude in $\NeqFour$
super-Yang-Mills theory.  Because the triangle and bubble diagrams
vanish, \eqn{FourFourNumeratorCrossing} is equivalent to the duality
relations (\ref{BCJDuality}). Thus, this representation of the
amplitude trivially satisfies the duality.

Using \eqn{OneloopGravityGen}, by replacing color factors with
numerators and compensating for the coupling change, we then
immediately have the four-point $\NeqEight$ supergravity amplitude,
\begin{equation}
{\cal M}_{\NeqEight}^{\oneloop}(1,2,3,4)= -\Bigl(\frac{\kappa}{2} \Bigr)^4
   [st A^\tree(1,2,3,4)]^2
\Bigl(I_4^{1234} + 
      I_4^{1243} +
      I_4^{1423} \Bigr) \,,
\label{NeqEightGravityFourPoint}
\end{equation}
which matches the known amplitude~\cite{GSB,BDDPR}.

We now generalize to supergravity amplitudes with fewer
supersymmetries. Specifically, consider the one-loop four-graviton
amplitudes with the $\NeqSix$ and $\NeqFour$ matter multiplets in 
the loop. These multiplets can be expressed as products of two gauge-theory multiplets:
\begin{eqnarray}
&& \NeqSix\ \hbox{matter} :  (\NeqFour\ \hbox{sYM}) \times  (\NeqOne\ \hbox{sYM})_{\matter} \,, \nn \\
&& \NeqFour\ \hbox{matter} : 
   (\NeqFour\ \hbox{sYM}) \times  (\hbox{scalar}) \,,
\end{eqnarray}
where the $\NeqOne$ Yang-Mills matter multiplet consists of a Weyl
fermion with two real scalars (this combination actually has two-fold supersymmetry so it can also be thought of as a $\NeqTwo$ matter multiplet), and on the second line ``(scalar)" denotes a single real scalar. 

Following \eqn{OneloopGravityGen}, we get the gravity amplitude by
taking the first copy of the gauge-theory amplitude and 
replacing the color factors with the kinematic numerator of the second
copy, constrained to satisfy the duality (\ref{BCJDuality}), and
switching the coupling to the gravitational one.  Because the duality
satisfying $\NeqFour$ super-Yang-Mills kinematic factors at four
points (\ref{FourFourNumeratorCrossing}) are independent of the loop
momentum, they simply come out of the integral as in
\eqn{SimpleOneloopGravitySusy} and behave essentially the same way as
color factors. Thus, we have a remarkably simple general formula at
four points,
\begin{eqnarray} 
 \mathcal{M}_{ {\cal N}+4\,\,{\rm susy}}^{\text{1-loop}} (1,2,3,4) &=&
\Bigl( \frac{\kappa}{2} \Bigr)^4 i s t A^{\text{tree}} (1, 2, 3, 4)  
 \Bigl(A_{ {\cal N}\,{\rm susy}}^{\text{1-loop}} (1, 2, 3, 4)  
 +  A_{ {\cal N}\,{\rm susy}}^{\text{1-loop}} (1, 2, 4, 3) \nn \\
&& \hskip 4.5 cm \null
 +  A_{ {\cal N}\,{\rm susy}}^{\text{1-loop}} (1, 4, 2, 3)\Bigr)\,,
\label{NeqM}
 \end{eqnarray}
where $A_{ {\cal N}\,{\rm susy}}^{\text{1-loop}}$ are one-loop color-
and coupling-stripped gauge-theory amplitudes for a theory with
${\cal N}$ (including zero) supersymmetries.  We were able 
pull out an overall $s t A^\tree  (1, 2, 3, 4)$ because of
the crossing symmetry apparent in \eqn{FourFourNumeratorCrossing}.

Using \eqn{NeqM} we can straightforwardly write down the four-graviton supergravity
amplitude $\mathcal{M}_{\NeqSix, {\matter}}^{\text{1-loop}}(1^{-},2^{-},3^{+},4^{+})$ with the $\NeqSix$ matter multiplet in the loop.  We use the
$\NeqOne$ one-loop amplitude representation\footnote{
Here we removed the factor of $i (-1)^{m+1} (4\pi)^{2-\eps}$ 
present in the integrals of ref.~\cite{BernMorgan}, where $m$ is
  2 for the bubble, 3 for the triangle and 4 for the box. (Compare eq.~(\ref{mPtsIntegral}) with  eq.~(A.13) of ref.~\cite{BernMorgan}.)}  from
ref.~\cite{BernMorgan} which is valid to all order in the dimensional
regularization parameter $\epsilon$:
\begin{eqnarray}
A_{\mathcal{N}=1, {\matter}}^{\oneloop} (1^-,  2^-, 3^+, 4^+) &=&
  i g^4  A^{\tree}(1^-,  2^-,3^+, 4^+) \Big(t J_4(s,t) - I_2(t)\Big)
    \,, \nn \\
A_{\mathcal{N}=1, {\matter}}^{\oneloop} (1^-,  2^-, 4^+, 3^+) &=&
  i g^4  A^{\tree}(1^-,  2^-,3^+, 4^+) \Big(t J_4(s,u) - \frac{t}{u}I_2(u)\Big)
    \,, \nn \\
A_{\mathcal{N}=1, {\matter}}^\oneloop (1^-,  4^+, 2^-, 3^+) &=&
 i g^4 A^\tree(1^-,  2^-, 3^+, 4^+) \Bigl(
   I_2(t)  + { t \over u } I_2(u) \nn \\
&& \null \hskip 2cm 
- t J_4(t,u) 
  - t I_4^{D=6-2\eps}(t,u) 
\Bigr) \, ,
\end{eqnarray}
where the integrals $I_2, J_4$ and $I_4^{D=6-2\eps}$ are defined in 
\app{IntegralsAppendix}. Using \eqn{NeqM} we
can see that the bubble integrals cancel and we have the amplitude in a
form valid to all orders in $\epsilon$. Also using the relation $J_4 = -\eps I_4^{D=6-2\eps}$, we get
\begin{eqnarray}
\mathcal{M}_{\NeqSix, \matter}^{\text{1-loop}}(1^{-},2^{-}, 3^{+},4^{+})
&=&\Bigl(\frac{\kappa}{2} \Bigr)^4 \frac{1}{s}
 [st A^\tree(1^-,  2^-,3^+, 4^+)]^2 \\
&& \null \times \Bigl[I_4^{D=6-2\epsilon}(t,u) 
 + \epsilon \Bigl( -I_4^{D=6-2\epsilon}(t,u) + I_4^{D=6-2\epsilon}(s,t) \nn\\
&& \hskip 4cm  \null 
   + I_4^{D=6-2\epsilon}(s,u)\Bigr) \Bigr]\,.\nn
\end{eqnarray}
Using the explicit value of $I_4^{D=6-2\eps}$ given in \eqn{6DboxEps}, 
we get the remarkably simple result to order $\epsilon^0$,
\begin{eqnarray}
\mathcal{M}_{\NeqSix, \matter}^{\text{1-loop}}(1^{-},2^{-}, 3^{+},4^{+})
&=&\frac{i c_{\Gamma}}{2} \Bigl(\frac{\kappa}{2} \Bigr)^4
   [st A^\tree(1^-,  2^-,3^+, 4^+)]^2  \: \frac{1}{s^2}   \left[\text{ln}^2\left(\frac{-t}{-u}\right) + \pi^2\right]  + \Ord(\eps)   \nn\\
&=& -\frac{i c_{\Gamma}}{2} \Bigl(\frac{\kappa}{2} \Bigr)^4
\frac{\spa{1}.{2}^4\spb{3}.{4}^4} {s^2} 
 \left[\text{ln}^2\left(\frac{-t}{-u}\right) + \pi^2\right] + \Ord(\eps)\,,
 \hskip .5 cm 
\end{eqnarray}
where the constant $c_{\Gamma}$ is defined in \eqn{cgamma}.
On the last line we plugged in the value of the tree amplitude, $st
A^\tree(1^-,2^-,3^+,4^+)=-i \spa{1}.{2}^2\spb{3}.{4}^2$.  Indeed, this
reproduces the known result from ref.~\cite{DunbarNorridge}.

Now consider the four-graviton amplitude with an $\NeqFour$
supergravity matter multiplet going around the loop.  We take the
four-gluon amplitudes with a scalar in the loop from
ref.~\cite{BernMorgan}. These are 
\begin{eqnarray}
A^\oneloop_{\scalar}(1^-, 2^-, 3^+, 4^+) \!& = &\!
  -ig^4A^{\rm tree}(1^-, 2^-, 3^+, 4^+)
  \Bigl( {1 \over t} I_2^{D=6-2\eps}(t) 
  + {1 \over s}  J_2(t) 
  - {t \over s} K_4 (s,t) \Bigr) \,, \nn \\
A^\oneloop_{\scalar}(1^-, 2^-, 4^+, 3^+) \!& = &\!
  -ig^4A^{\rm tree}(1^-, 2^-, 3^+, 4^+)
  \Bigl( {t \over u^2} I_2^{D=6-2\eps}(u) 
  + {t \over su}  J_2(u) 
  - {t \over s} K_4 (s,u) \Bigr) \,, \nn \\
A^\oneloop_{\scalar} (1^-, 4^+, 2^-, 3^+)\! &=&\!
  -i g^4 A^{\rm tree}(1^-, 2^-, 3^+, 4^+) \biggl(\!
       -{ t(t-u) \over s^2 } J_3(u)
      -{  t(u-t) \over s^2 } J_3(t)
      -{ t^2 \over s^2 } I_2(u)\nn \\
      & & \null
      -{ t u \over s^2 } I_2(t) 
      -{ t \over u^2 } I_2^{D=6-2\eps}(u)
      -{ 1 \over t } I_2^{D=6-2\eps}(t)
      -{ t \over su } J_2(u)
      -{ 1 \over s } J_2(t) \nn \\
     & & \null 
      +{ t \over s } I_3^{D=6-2\eps}(u)
      +{ t \over s } I_3^{D=6-2\eps}(t)
      +{ t^2 u \over s^2 } I_4^{D=6-2\eps} (t,u)
      -{  t \over s } K_4 (t,u)\biggr) \,, \nn \\
\end{eqnarray}
where the integral functions are given in \app{IntegralsAppendix}.
Using \eqn{NeqM}, we immediately have a form for the 
contributions of an $\NeqFour$ supergravity matter multiplet
valid to all orders in $\epsilon$, 
\begin{eqnarray}
\mathcal{M}_{\NeqFour, \matter}^{\text{1-loop}}(1^-, 2^-, 3^+, 4^+)&=&
 \Bigl(\frac{\kappa}{2} \Bigr)^4 [st A^\tree(1^-,  2^-,3^+, 4^+)]^2 \biggl(
       -{ (t-u) \over s^3 } J_3(u)
      -{  (u-t) \over s^3 } J_3(t)\nn \\
      & &\null
      -{ t \over s^3 } I_2(u)
      -{ u \over s^3 } I_2(t) 
      +{ 1 \over s^2 } I_3^{D=6-2\eps}(u)
      +{ 1 \over s^2} I_3^{D=6-2\eps}(t)
        \\
      & & \null
      +{ t u \over s^3 } I_4^{D=6-2\eps} (t,u)
      -{  1 \over s^2 } K_4 (t,u) -{1 \over s^2 } K_4 (s,t) -{  1 \over s^2 } K_4 (s,u)\biggr) \nn \,.
\end{eqnarray}
Expanding this through order $\epsilon^0$ and using
integral identities from refs.~\cite{BernMorgan,BrandhuberLoops} (see
also \app{IntegralsAppendix}) to reexpress everything in terms of
six-dimensional boxes, bubbles and rational terms, we obtain
\begin{eqnarray}
\mathcal{M}_{\NeqFour, \matter}^{\text{1-loop}}(1^-, 2^-, 3^+,4^+) 
&=& \frac{1}{2}\Bigl(\frac{\kappa}{2} \Bigr)^4
\frac{\spa{1}.{2}^2\spb{3}.{4}^2}{\spb{1}.{2}^2\spa{3}.{4}^2}\Big[ i
  c_{\Gamma} s^2 + s (u - t)\Big(I_2(t)-I_2(u)\Big)\nn \\
&& \null  \hskip 3 cm 
 -2 I_4^{D=6-2\eps}(t,u) s t u\Big] 
+ \Ord(\eps) \,, \hskip 2 cm 
 \end{eqnarray}
matching the result of ref.~\cite{DunbarNorridge}. 

\subsection{Five-point one-loop ${\cal N} \ge 4$ supergravity amplitudes}
\label{FivePtsGravSubsection}

Our construction at five points is again directly based on
\eqn{OneloopGravityGen}. We only need to construct ${\cal
M}_{{\cal N} \rm susy}^{\oneloop}(1^-,2^-,3^+,4^+,5^+)$; 
the other nonvanishing amplitudes are related by parity and relabeling.
%
%
\begin{figure*}[tb] 
\begin{center}
\includegraphics[scale=0.58]{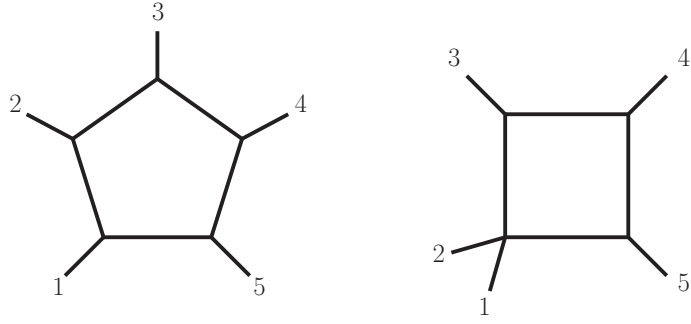}
\end{center}
\vskip -.9 cm 
\caption[a]{\small Pentagon and box integrals appearing in the 
$\NeqFour$ super-Yang-Mills five-point one-loop amplitudes.  The complete set
of such integrals
is generated by permuting external legs and removing overcounts.}
\label{PentagonBoxFigure}
\end{figure*}
%
%
Our starting point is the known one-loop
five-point amplitudes of $\NeqFour$ super-Yang-Mills theory.  The
original construction of the amplitude~\cite{FiveYM,UnitarityMethodI} uses a
basis of scalar box integrals.  Rearranging these results into 
the adjoint-representation color basis gives
\begin{equation}
{\cal A}^{\text{1-loop}}(1,2,3,4,5) = g^5 
\sum_{S_5/(Z_5\times Z_2)} \!\! c_{12345}\, A^{\text{1-loop}}(1,2,3,4,5)\,.
\label{NeqFourYMFourPt}
\end{equation}
The sum runs over the distinct permutations of the external legs of
the amplitude.  This is the set of all $5!$ permutations, $S_5$, but
with cyclic, $Z_5$, and reflection symmetries, $Z_2$, removed, leaving
12 distinct permutations.  The color factor $c_{12345}$ is the one of
the pentagon diagram shown in \fig{PentagonBoxFigure}, with legs
following the cyclic ordering as in \eqn{AdjointTrace}.  The
color-ordered one-loop amplitudes of $\NeqFour$ super-Yang-Mills
theory are
\begin{eqnarray}
A_{\NeqFour}^{\oneloop}(1,2,3,4,5) &=&\frac{i}{2}A^\tree(1,2,3,4,5)
\Bigl(s_{34}s_{45} I_4^{(12)345}+s_{45}s_{15} I_4^{1(23)45}+
s_{12}s_{15} I_4^{12(34)5}\nn \\
&&\hskip3.3cm \null
+s_{12}s_{23} I_4^{123(45)}+s_{23}s_{34} I_4^{234(51)} \Bigr) + \Ord(\eps)\,,
\hskip 1 cm 
\label{FiveMHV}
\end{eqnarray}
where $s_{ij} = (k_i + k_j)^2$ and the 
 $I_4^{abc(de)}$ are box integrals where the legs in parenthesis
connects to the same vertex, {\it e.g.} $I_4^{(12)345}$ is the box
diagram in \fig{PentagonBoxFigure}.  The explicit value of
$I_4^{(12)345}$ is given in \eqn{OneMassBoxIntegral}, and the values
of the remaining box integrals are obtained by relabeling.  If we
insert these explicit expressions in \eqn{FiveMHV} then the
polylogarithms cancel after using identities (see
refs.~\cite{FiveYM,UnitarityMethodI}) leaving the expression for
$A^{\text{1-loop}}_{\NeqFour}$ given in \eqn{ExplicitFivePt} of
\app{FivePointAppendix}. The representation (\ref{FiveMHV}) of the amplitude does not manifestly satisfy the duality. 

A duality satisfying representation of the amplitude was found in
ref.~\cite{fivepointBCJ}:
\begin{eqnarray} 
\mathcal{A}_{\NeqFour}^{\text{1-loop}}(1^-,2^-,3^+,4^+,5^+)& =& g^5 \spa{1}.{2}^4
\Big(\sum_{S_5/(Z_5\times Z_2)}\hskip -.2 cm 
 c_{12345} n_{12345}  I_5^{12345} \nn \\
&& \null \hskip 2 cm 
  + \sum_{S_5/Z_2^2} c_{[12] 345} n_{[12]345} \frac{1}{s_{12}}I_4^{(12)345} \Big)
\,,
\hskip 1 cm 
\label{FiveMHVAlt}
\end{eqnarray}
where $I_5^{12345}$ is the scalar pentagon, and $I_4^{(12)345}$ is the
one-mass scalar box integral, as shown in \fig{PentagonBoxFigure}.
The explicit values of these integrals through $\Ord(\eps^0)$ are
collected in \app{IntegralsAppendix}. Each of the two sums runs over
the distinct permutations of the external legs of the integrals.  For
$I_5^{12345}$, the set $S_5/(Z_5\times Z_2)$ denotes all permutations
but with cyclic and reflection symmetries removed, leaving 12 distinct
permutations.  For $I_4^{(12)345}$ the set $S_5/Z_2^2$ denotes all
permutations but with the two symmetries of the one-mass box removed,
leaving 30 distinct permutations.  Note that we pulled out an overall
factor $\spa{1}.{2}^4$, which we do not include in the numerators. (If
promoted to its supersymmetric form it should then be
included~\cite{fivepointBCJ}.)  The numerators defined in this way are
then~\cite{fivepointBCJ}
\begin{equation}
n_{12345} = - \frac{ \spb1.2 \spb2.3\spb3.4 \spb4.5 \spb5.1}
 {4 i \epsilon(1,2,3,4)}\,,
\label{pentnumer}
\end{equation}
and
\begin{equation}
n_{[12]345} = \frac{\spb1.2^2 \spb3.4 \spb4.5 \spb5.3} 
  {4 i \epsilon(1,2,3,4)}\,,
\label{boxnumer}
 \end{equation}
where $ 4 i \epsilon(1,2,3,4)=4 i \epsilon_{\mu \nu \rho \sigma}k_1^{\mu}k_2^{\nu}k_3^{\rho}k_4^{\sigma}=\spb{1}.{2}\spa{2}.{3}\spb{3}.{4}\spa{4}.{1}-\spa{1}.{2}\spb{2}.{3}\spa{3}.{4}\spb{4}.{1}$.
It is not difficult to confirm that the duality holds for this representation,
for example,
\begin{equation}
n_{12345} - n_{21345} = n_{[12]345}\,.
\end{equation}
A nice feature of this representation is that the numerator factors of
both the pentagon and box integrals do not depend on loop momentum,
allowing us to use \eqn{SimpleOneloopGravitySusy}. This will greatly
simplify the construction of the corresponding supergravity amplitudes.

We first consider the one-loop five-point $\NeqEight$ amplitude.  In
this case we have several useful representations.  Proceeding as in
\sect{FourPtsGravSubsection}, using \eqn{SimpleOneloopGravitySusy}, 
we can obtain the
five-point amplitude for $\NeqEight$ by replacing the color factors in
\eqn{NeqFourYMFourPt} with the numerator factors of \eqn{pentnumer},
multiplying by the overall factor $\spa{1}.{2}^4$, and putting in the
gravitational couplings. This yields
\begin{eqnarray}
{\cal M}^{\text{1-loop}}_{\NeqEight}(1^-, 2^-, 3^+, 4^+, 5^+) &=& \frac{i}{2}
\Bigl({\kappa \over 2} \Bigr)^5\spa{1}.{2}^4 
 \sum_{S_5/Z_2}  n_{12345}  A^\tree(1^-, 2^-, 3^+, 4^+, 5^+) s_{12}s_{23} 
     I_4^{123(45)} \nn \\
&& \hskip 4 cm \null + \Ord(\eps) \,,
\label{FiveGrav}
\end{eqnarray}
where the sum runs over all permutations of external legs, denoted 
by $S_5$, but with reflections $Z_2$ removed. 
To obtain a second representation, we can instead replace the color
factors in \eqn{FiveMHVAlt} with their corresponding numerator
factors, yielding an alternative expression for the amplitude,
\begin{equation} 
{\cal M}^{\text{1-loop}}_{\NeqEight}(1^-, 2^-, 3^+, 4^+, 5^+) =  
\Bigl({\kappa \over 2} \Bigr)^5 \spa{1}.{2}^8  
\Big(\sum_{S_5/(Z_5\times Z_2)} \hskip -.3 cm  (n_{12345})^2  I_5^{12345} + 
   \sum_{S_5/Z_2^2} (n_{[12]345})^2 \frac{1}{s_{12}}I_4^{(12)345} \Big) \,,
\label{FiveGravAlt}
\end{equation}
where the sums run over the same permutations as in \eqn{FiveMHVAlt}.
We have checked that in $D=4$ both formulas (\ref{FiveGrav}) and
(\ref{FiveGravAlt}) are equivalent to the known five-point amplitude
from ref.~\cite{MHVOneLoopGravity} (after reducing the scalar pentagon
integrals to one-mass box integrals),
\be
\label{knownN8}
 \mathcal{M}^{\text{1-loop}}_{\NeqEight}(1^-, 2^-, 3^+, 4^+, 5^+) =
         \Bigl({\kappa \over 2} \Bigr)^5 \spa1.2^8\,
         \sum_{S_5/Z_2^2} d^{123(45)}_{\NeqEight}\, I_4^{123(45)} 
+\Ord(\eps)\,,
\ee
where the box coefficient is given by
\be
 d^{123(45)}_{\NeqEight}\equiv-{1\over 8}h(1,\{2\},3) h(3,\{4,5\},1) 
      \tr^2[\s k_1 \s k_2 \s k_3( \s k_4+ \s k_5)]\,, 
\label{N8BoxCoefficient}
\ee
and the ``half-soft'' functions are
\be
h(a,\{2\},b) \equiv \frac{1}{\spa a.2^2 \spa 2.b^2} \,, ~~~~~h(a,\{4,5\},b) \equiv \frac{\spb 4.5}{\spa 4.5  \spa a.4 \spa 4.b \spa a.5 \spa 5.b}\,.
\ee
Indeed it is straightforward to check that 
\be
\spa{1}.{2}^4d^{123(45)}_{\NeqEight}=\frac{i}{2} \,s_{12}s_{23} \Big(n_{12345} A^\tree(1^-, 2^-, 3^+, 4^+, 5^+)+n_{12354} A^\tree(1^-, 2^-, 3^+, 5^+, 4^+) \Big)\,,
\ee
where the pentagon numerator is given in \eqn{pentnumer}.

Let us now study amplitudes with fewer supersymmetries starting with the
five-graviton amplitude with the $\NeqSix$ matter multiplet running
around the loop.  We pick the helicities $(1^-,2^-,3^+, 4^+, 5^+)$ for
the gravitons; as noted above all other helicity or particle configurations
can be obtained from this.  For the $\NeqSix$ and $\NeqFour$
matter multiplets from \eqn{SimpleOneloopGravitySusy} we have
\begin{eqnarray}
M_{\NeqSix, \matter}(1^-,2^-,3^+,4^+,5^+) &=& \Bigl({\kappa \over 2} \Bigr)^5 \spa1.2^4 
\sum_{S_5/(Z_5\times Z_2)} \hskip -.3 cm 
n_{12345}\, A^\oneloop_{\NeqOne, \matter}(1^-,2^-,3^+,4^+,5^+)\,,\hskip 1 cm\nn \\
M_{\NeqFour, \matter}(1^-,2^-,3^+,4^+,5^+) &=& \Bigl({\kappa \over 2} \Bigr)^5\spa1.2^4 
\sum_{S_5/(Z_5\times Z_2)} \hskip -.3 cm 
n_{12345}\, A^\oneloop_{\scalar}(1^-,2^-,3^+,4^+,5^+)\,, \hskip 1cm 
\label{NeqFourSixMatter}
\end{eqnarray}
where $n_{12345}$ is given in \eqn{pentnumer} and the sums run over all
 permutations, but with cyclic ones and the reflection removed.

There are a number of simplifications that occur because of the
permutation sum in \eqn{NeqFourSixMatter} and because of the algebraic
properties of the $\NeqFour$ sYM numerators ($n_{12345}$ and
permutations).  Because the matter multiplet contributions have
neither infrared nor ultraviolet divergences~\cite{DunbarNorridgeInfinities}, 
all $1/\eps^2$ and $1/\eps$ divergences cancel.  In $\NeqSix$
supergravity, this manifests itself by the cancellation of all bubble
and triangle integral contributions, as noted in
ref.~\cite{DunbarSugra}.  In the case of $\NeqFour$ supergravity, the
cancellation is not complete but the sum over bubble-integral
coefficients vanishes to prevent the appearance of a $1/\eps$
singularity.  A rational function remains which can be written in a
relatively simple form once the terms are combined and simplified.
Our results match those obtained in ref.~\cite{DunbarSugra}.

The final form of the  $\NeqSix$ results after simplifications 
are then~\cite{DunbarSugra}
\begin{eqnarray}
\label{knownN6}
&& \mathcal{M}^{\text{1-loop}}_{\NeqSix,\matter}(1^-, 2^-, 3^+, 4^+, 5^+) 
 \nn\\
&& \null \hskip 1 cm 
= -   \Bigl({\kappa \over 2} \Bigr)^5 \spa1.2^8\,
         \sum_{Z_3(345)}
 \biggl(\frac{\spa1.3 \spa2.3 \spa1.4 \spa 2.4}
                            {\spa{3}.{4}^2 \spa1.2^2} \biggr) 
    \Bigl(d^{324(51)}_{\NeqEight} \, I_{4, \rm trunc}^{324(51)} +
          d^{314(52)}_{\NeqEight} \, I_{4,\rm trunc}^{314(52)} \Bigr) \nn\\
&& \hskip 4 cm \null 
+ \Ord(\eps) \,,
\end{eqnarray}
where the summation runs over the three cyclic permutations of legs 3,
4, 5 in the box integrals and coefficients. The factor 
$d^{123(45)}_{\NeqEight}$ is exactly the coefficient (\ref{N8BoxCoefficient})
of the $\NeqEight$ theory and the integral $I_{4,\rm
trunc}^{123(45)}$ given in \eqn{OneMassBoxIntegralTrunc} of
\app{IntegralsAppendix} is the one-mass box integral but with its infrared
divergent terms subtracted out. 
Similarly, the simplified $\NeqFour$ supergravity results are
\begin{eqnarray}
\label{knownN4}
&& \mathcal{M}^{\text{1-loop}}_{\NeqFour,\matter}(1^-, 2^-, 3^+, 4^+, 5^+) 
 \nn\\
&& \null \hskip 1 cm 
=    \Bigl({\kappa \over 2} \Bigr)^5 \biggl[\spa1.2^8\,
         \sum_{Z_3(345)}
 \biggl(\frac{\spa1.3 \spa2.3 \spa1.4 \spa 2.4}
                            {\spa{3}.{4}^2 \spa1.2^2} \biggr)^2 
    \Bigl(d^{324(51)}_{\NeqEight} \, I_{4, \rm trunc}^{324(51)} +
          d^{314(52)}_{\NeqEight} \, I_{4,\rm trunc}^{314(52)} \Bigr) \nn \\
&& \null \hskip 3 cm 
+ i c_\Gamma \sum_{i=3}^5 (c_{1i} \ln(-s_{1i}) + c_{2i} \ln(-s_{2i}) ) 
+ i c_\Gamma R_5 \biggr] + \Ord(\eps) \,,
\end{eqnarray}
where the coefficient of $\log(-s_{13})$ coming from 
the bubble integrals is
\begin{eqnarray} \label{DunBubble}
c_{13} &=& 
\frac{1}{2}\,
\frac{\spa1.2^4 \spb 3.1 \spb 5.2}{\spa1.3 \spa2.5\spa 4.5} \Bigg [ -\frac{ \spa 2.4^2  \langle 4 |2+5| 4 ] \ \langle 1|3| 4]^2}{\spa 3.4^2 \spa 4.5  \langle 4 |1+3| 4]^2} - \frac {\spa 2.3}{\spa 3.4} \Bigg ( \frac{\spa 1.5 \spa 2.5 \langle 1 |3| 5] \ \langle 5 |2| 4]} {\spa 3.5^2 \spa 4.5 \langle 5 |1+3| 5 ]} \nn \\
&&\null - \frac{\spa 1.4 \spa 2.4 \langle 1|3|4] \ \langle 4|2+5|4]}{ \spa 3.4^2 \spa 4.5 \langle 4 |1+3| 4]}\Bigg ) + \frac{\spa 2.4}{\spa 3.4} \Bigg ( \frac{\spa 1.4 \spa 2.3 \langle 1|3|4] \ \langle 3 |2+5| 4]}{\spa 3.4^2 \spa 3.5 \langle 4 |1+3| 4]}   \nn \\
&&\null + \frac{ \spa 2.5 \langle 5|2|4]}{\spa 3.5 \spa 4.5} \Bigg ( \frac{ \spa 1.5 \langle 1|3|5]}{\spa 4.5 \langle 5|1+3|5]} - \frac{\spa 1.4 \langle 1|3|4]}{\spa 4.5 \langle 4|1+3|4]} \Bigg ) \Bigg) \Bigg ] \, + ( 4 \leftrightarrow 5)\,,
\end{eqnarray}
and the others are given by the natural label swaps,
$c_{1i}=c_{13}|_{3\leftrightarrow i}$ and
$c_{2i}=c_{1i}|_{1\leftrightarrow 2}$.  The rational terms follow
the notation of ref.~\cite{DunbarSugra},
\begin{equation}
R_5 =   R_5^b\; +\!\! \sum_{Z_2(12)\times Z_3(345)} R_5^a \,,
\label{Rsum}
\end{equation}
where 
\begin{eqnarray}
R_5^a &=& - \frac{1}{2} \spa 1.2^4 \frac{ \spb 3.4^2 \spb 2.5 \spa 2.3 \spa 2.4}{\spa 3.4^2 \spa 2.5 \spa 3.5 \spa 4.5}\,,  \hskip 1.5 cm 
R_5^b = -  \spa 1.2^4 \frac{ \spb 3.4 \spb 3.5 \spb 4.5}{\spa 3.4 \spa 3.5 \spa 4.5}\,. \hskip .5 cm 
\end{eqnarray}
The sum in \eqn{Rsum} corresponds to the composition of the two permutations of
negative-helicity legs 1 and 2 and the three cyclic permutations over the
positive-helicity legs 3, 4 and 5, giving six terms in total.
(Results for general MHV amplitudes may be found in
ref.~\cite{DunbarSugra}.)

Inserting the results from \eqn{NeqFourSixMatter} into
\eqn{SugraAssembly} immediately converts the results we obtained for
the matter multiplets into those for the ${\cal N} = 4,5,6$ gravity
multiplets (the pure supergravities).  For the $\NeqFour$ and
$\NeqSix$ gravity multiplets these match the results of
ref.~\cite{DunbarSugra}.

Thus we have succeeded in expressing the four- and five-point
integrated amplitudes of ${\cal N} \ge 4$ supergravity amplitudes as
simple linear combination of corresponding gauge-theory ones.  To
generalize this construction to higher points, one would need to find duality
satisfying representations of $m$-point one-loop $\NeqFour$ super-Yang-Mills
amplitudes.

\subsection{Comments on two loops}

\begin{figure*}[tb] 
\begin{center}
\includegraphics[scale=0.45]{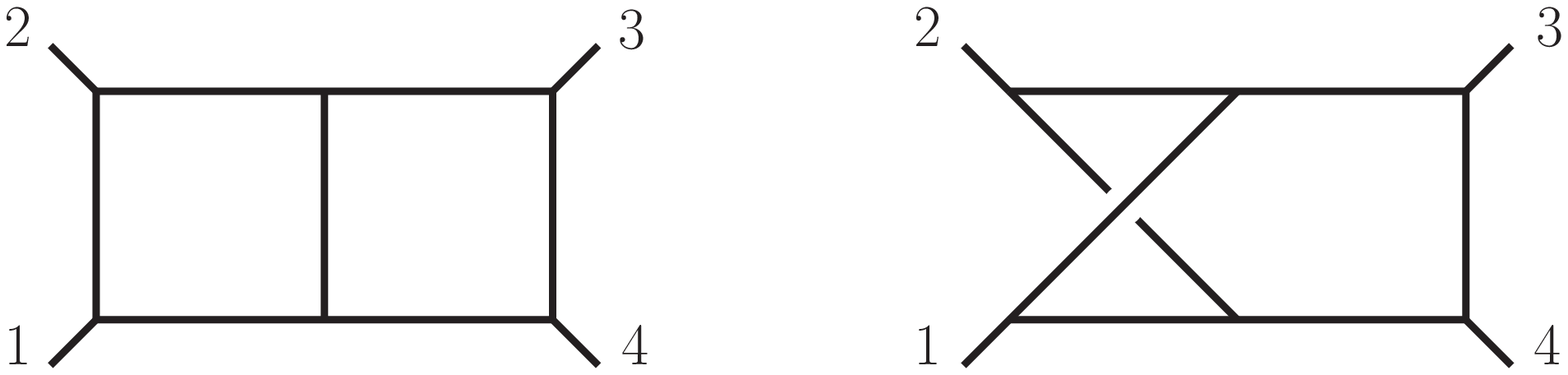}
\end{center}
\vskip -.9 cm 
\caption[a]{\small The two-loop cubic diagrams appearing in the
two-loop four-point $\NeqFour$  and $\NeqEight$ supergravity amplitudes.
}
\label{TwoloopDiagramFigure}
\end{figure*}

An interesting question is whether the same considerations hold at
higher loops.  Consider the two-loop four-point amplitude of
$\NeqFour$ super-Yang-Mills theory~\cite{BRY,BDDPR}:
\begin{eqnarray}
{\cal A}_4^{\twoloop}(1,2,3,4) &=&  - g^6 st  \,
              A_4^{\rm tree}(1,2,3,4) \Bigl(
    c^{\P}_{1234} \, s \, I_4^{\twoloop,\P}(s, t)
  + c^{\P}_{3421} \, s \, I_4^{\twoloop,\P}(s, u)
\label{TwoLoopYM} \\
 && \null \hskip  2.5 truecm
  + c^{\NP}_{1234} \, s\, I_4^{\twoloop , \NP}(s,t)
  + c^{\NP}_{3421} \, s\, I_4^{\twoloop , \NP}(s,u)
+  {\rm cyclic} \Bigr)\,, \nn
\end{eqnarray}
where `$+$~cyclic' instructs one to add the two cyclic permutations of
(2,3,4) and the integrals correspond to the scalar planar and 
nonplanar double-box diagrams displayed in \fig{TwoloopDiagramFigure}.
As at one loop, the color factor for each diagram is obtained by
dressing each cubic vertex with an $\f^{abc}$.  It is then simple to check
that all duality relations (\ref{BCJDuality}) hold.

According to the double-copy prescription (\ref{DoubleCopy}), we
obtain the corresponding $\NeqEight$ supergravity amplitude
by replacing the color factor with a numerator factor,
\begin{equation}
 c^{\P}_{1234} \rightarrow  i s^2 t A^\tree(1,2,3,4)\,,  \hskip 2 cm 
 c^{\NP}_{1234} \rightarrow i s^2 t A^\tree(1,2,3,4) \,,
\label{ColorReplacement}
\end{equation}
including relabelings and then swapping the gauge coupling for the
gravitational one.  Indeed, this gives the correct $\NeqEight$
supergravity amplitude, as already noted in ref.~\cite{BDDPR}.

As explained in \sect{ReviewSection}, generalized gauge invariance
implies that we need have only one of the two copies in a form
manifestly satisfying the duality (\ref{BCJDuality}). The color Jacobi
identity allows us to express any four-point color factor of an
adjoint representation in terms of the ones in
\fig{TwoloopDiagramFigure}~\cite{DixonColor}. If the duality and
double-copy properties hold we should then be able to obtain
integrated ${\cal N} \ge 4$ supergravity amplitudes starting from
${\cal N} \le 4$ super-Yang-Mills theory and applying the replacement rule
(\ref{ColorReplacement}).  Indeed, in ref.~\cite{CamilleForthcoming},
explicit expressions for the four-point two-loop ${\cal N} \ge 4$
supergravity amplitudes, including the finite terms, are obtained in
this manner.

Two-loop supergravity amplitudes are UV finite and their IR behavior
is given in terms of the square of the one-loop amplitude
\cite{SchnitzerIR}:
\begin{equation}\label{IRexp}
\mathcal{M}_4^{(\text{2-loop})}(\epsilon)/\mathcal{M}_4^{\tree} = 
\frac{1}{2} \Big [ \mathcal{M}_4^{\text{1-loop}} (\epsilon)/
        \mathcal{M}_4^\tree \Big ]^2 
 + \text{finite} \, .
\end{equation}
The amplitudes of ref.~\cite{CamilleForthcoming} satisfy this relation
and the finite remainders are given in a relatively simple
form. These two-loop results then provide a rather nontrivial confirmation
of the duality and double-copy properties for cases with less than
maximal supersymmetry.

\section{Conclusions}
\label{ConclusionSection}

The duality between color and kinematic numerators offers a powerful
means for obtaining loop-level gauge and gravity amplitudes and for
understanding their structure. A consequence of the duality conjecture
is that complete amplitudes are controlled by a set of master
diagrams; once the numerators are known in a form that makes the
duality between color and kinematics manifest, all others are
determined from Jacobi-like relations.  In this form we immediately
obtain gravity integrands via the double-copy relation.  

In the present paper, we used the duality to find examples where
{\it integrated} supergravity amplitudes are expressed directly as linear
combinations of gauge-theory amplitudes.  In particular, we
constructed the integrated four- and five-point one-loop amplitudes of
${\cal N} \ge 4$ supergravity directly from known gauge-theory
amplitudes.  This construction was based on identifying
representations of $\NeqFour$ super-Yang-Mills four- and five-point
amplitudes that satisfy the duality.  Because the relations are valid
in $D$ dimensions, by using known $D$-dimensional forms of
gauge-theory four-point amplitudes we obtain corresponding ones for
supergravity.  The agreement of our four- and five-point ${\cal N} \ge
4$ supergravity results with independent
evaluations~\cite{DunbarNorridge, DunbarSugra} in $D=4$ provides
evidence in favor of these conjectures holding for less than maximal
supersymmetry. The two-loop results in ref.~\cite{CamilleForthcoming}
provide further nontrivial evidence.

The examples we presented here are particularly simple because the
numerator factors of one copy of the gauge-theory amplitudes were
independent of loop momenta.  In more general cases, we expect useful
constraints to arise at the integrated level.  These constraints, for
example, lead to KLT-like relations visible in box-integral
coefficients, such as those found in
refs.~\cite{MHVOneLoopGravity,Bjerrum2}.  It would be very interesting
to further explore relations between gravity and gauge theory after
having carried out the loop integration.

There are a number of other interesting related problems.  It would of
course be important to unravel the underlying group-theoretic
structure responsible for the duality between color and kinematics.
Some interesting progress has recently made for self-dual field
configurations and for MHV tree amplitudes, identifying an underlying
diffeomorphism Lie algebra~\cite{OConnell}.  Another key problem is to
find better means for finding representations that automatically
satisfy the duality and double-copy properties.  Such general
representations are known at tree level for any choice of
helicities~\cite{TreeAllN}. We would like to have similar
constructions at loop level, instead of having to find duality
satisfying forms case by case.  In particular, no examples have as yet
been constructed at loop level at six and higher points.

In summary, using the duality between color and kinematics we exposed
a surprising relation between integrated four- and five-point one-loop
amplitudes of ${\cal N} \ge 4$ supergravity and those of gauge theory.
We look forward to applying these ideas to further unravel the
structure of gauge and gravity loop amplitudes.

\section*{Acknowledgments}
\vskip -.3 cm
We especially thank Harald Ita for crucial discussions
developing the basic observations of this paper. We thank
Tristan Dennen and Lance Dixon for many stimulating discussions and for collaboration on related topics. We
also thank David Dunbar for assistance in comparing our results to 
those of ref.~\cite{DunbarSugra}.  This research was supported in part
by the US Department of Energy under contracts DE--AC02--76SF00515 and
DE-FG02-90ER40577 and by the National Science Foundation
under Grant No. NSF PHY05-51164.  HJ's research is supported by the
European Research Council under Advanced Investigator Grant
ERC-AdG-228301. CBV is also supported by a postgraduate scholarship
from the Natural Sciences and Engineering Research Council of Canada.
We thank the Kavli Institute for Theoretical Physics at Santa Barbara
for hospitality while this work was completed.

\appendix

\section{The one-loop five-point Yang-Mills amplitudes}
\label{FivePointAppendix}

This appendix collects the five-point one-loop Yang-Mills amplitudes
used to construct the five-point supergravity amplitudes. The external
states are gluons and all amplitudes can be obtained from two 
configurations, $(1^{-},2^{-},3^{+},4^{+},5^{+})$ and
$(1^{-},2^{+},3^{-},4^{+},5^{+})$, using relabeling and parity. These
results are from ref.~\cite{FiveYM} which the reader is invited to 
consult for further details.  The results are presented in the
four-dimension helicity (FDH) regularization scheme~\cite{FDH}, which
is known to preserve supersymmetry at one loop.

The five-gluon color-ordered and coupling-stripped amplitudes with the
$\NeqFour$, $\NeqOne$ matter multiplet and a {\it real} scalar going
around the loop can be expressed as:
\begin{eqnarray}
A^{\text{1-loop}}_{\NeqFour}(1,2,3,4,5) &=& c_{\Gamma} V^{g} A_5^{\tree} \,, \nn \\
A^{\text{1-loop}}_{\NeqOne, \matter}(1,2,3,4,5) &=& 
  -c_{\Gamma} ( V^{f} A_5^{\tree} + i F^{f} ) \,, \nn \\
A^{\text{1-loop}}_{\text{scalar}}(1,2,3,4,5) &=& {1\over 2} 
c_{\Gamma} (V^{s} A_5^{\tree} + i F^{s} ) \,, 
\label{ExplicitFivePt}
\end{eqnarray}
where the tree amplitudes are 
\begin{eqnarray}
\Atree_5(1^{-},2^{-},3^{+},4^{+},5^{+}) &= &  
\frac{i{\spa1.2}^4}{\spa1.2\spa2.3\spa3.4\spa4.5\spa5.1} \,, \nn \\
\Atree_5(1^{-},2^{+},3^{-},4^{+},5^{+})
 & =&  \frac{i{\spa1.3}^4}{\spa1.2\spa2.3\spa3.4\spa4.5\spa5.1}\,. 
\end{eqnarray}
The function, 
\begin{eqnarray}
\text{V}^{g} &=& -{1\over\e^2}\sum_{j=1}^5 (-s_{j,j+1})^{-\e}
          +\sum_{j=1}^5 \ln\L{-s_{j,j+1}\over -s_{j+1,j+2}}\R\,
                        \ln\L{-s_{j+2,j-2}\over -s_{j-2,j-1}}\R
          +{5\over6}\pi^2\,.
\end{eqnarray}
is independent of the helicity configuration.   In contrast to 
ref.~\cite{FiveYM}, we have set the dimensional-regularization scale 
parameter, $\mu$, to unity.
For the $(1^{-},2^{-},3^{+},4^{+},5^{+})$
helicity configuration we have,
\begin{eqnarray}
V^{f}&=&-{1\over\e}+{1\over2}\LB\ln\L{-s_{23}}\R
                                 +\ln\L{-s_{51}}\R\RB-2\,,
\hskip 10mm
V^{s} = -{1\over3} V^{f} + {2\over9}\,,\cr
F^{f}&=&-{1\over 2}
   {{\spa1.2}^2 \L\spa2.3\spb3.4\spa4.1+\spa2.4\spb4.5\spa5.1\R\over
    \spa2.3\spa3.4\spa4.5\spa5.1}
     {\Ll_0\L {-s_{23}\over -s_{51}}\R\over s_{51}}\,,\cr
F^{s}&=&-{1\over 3}
   {\spb3.4\spa4.1\spa2.4\spb4.5
     \L\spa2.3\spb3.4\spa4.1+\spa2.4\spb4.5\spa5.1\R\over\spa3.4\spa4.5}
     {\Ll_2\L {-s_{23}\over -s_{51}}\R\over s_{51}^3}
     - {1\over3}F^{f} \label{FiveAmplsA}\\
  \hskip 3mm &
   &\null -{1\over3}{\spa3.5{\spb3.5}^3\over\spb1.2\spb2.3\spa3.4\spa4.5\spb5.1}
     +{1\over3}{\spa1.2{\spb3.5}^2\over\spb2.3\spa3.4\spa4.5\spb5.1}
     +{1\over6}{\spa1.2\spb3.4\spa4.1\spa2.4\spb4.5\over
                  s_{23}\spa3.4\spa4.5 s_{51}}\,,\nn
\end{eqnarray}
and the corresponding functions for the $(1^{-},2^{+},3^{-},4^{+},5^{+})$
helicity configuration,
\begin{eqnarray}
V^{f}&=& -{1\over\e}+{1\over2}\LB\ln\L{-s_{34}}\R
                                 +\ln\L{-s_{51}}\R\RB-2\,,
\hskip 10mm
V^{s} = -{1\over3} V^{f} +{2\over9}\,,\cr
F^{f}&=&
      -{{\spa1.3}^2 {\spa4.1} {\spb2.4}^2
         \over {\spa4.5} {\spa5.1}}
           {\Ls_1\L {-s_{23}\over -s_{51}},\,{-s_{34}\over -s_{51}}\R
            \over s_{51}^2}
      +{{\spa1.3}^2 {\spa5.3} {\spb2.5}^2
         \over {\spa3.4} {\spa4.5}}
          {\Ls_1\L {-s_{12}\over -s_{34}},\,{-s_{51}\over -s_{34}}\R
           \over s_{34}^2}\cr
      &&\null -{1\over2} {{\spa1.3}^3
          (\spa1.5 \spb5.2 \spa2.3-\spa3.4 \spb4.2 \spa2.1)
             \over \spa1.2\spa2.3\spa3.4\spa4.5\spa5.1}
           {\Ll_0\L {-s_{34}\over -s_{51}}\R\over s_{51}}\,,\cr
F^{s} &=&
      - {{\spa1.2} {\spa2.3} {\spa3.4}
          {\spa4.1}^2 {\spb2.4}^2
         \over {\spa4.5} {\spa5.1} {\spa2.4}^2}\,
          {2 \, \Ls_1\L {-s_{23}\over -s_{51}},\,{-s_{34}\over -s_{51}}\R
           + \Ll_1\L {-s_{23}\over -s_{51}}\R
           + \Ll_1\L {-s_{34}\over -s_{51}}\R  \over s_{51}^2} \cr
      &&\null + {{\spa3.2} {\spa2.1} {\spa1.5}
          {\spa5.3}^2 {\spb2.5}^2
         \over {\spa5.4} {\spa4.3} {\spa2.5}^2}\,
          {2 \, \Ls_1\L {-s_{12}\over -s_{34}},\,{-s_{51}\over -s_{34}}\R
           + \Ll_1\L {-s_{12}\over -s_{34}}\R
           + \Ll_1\L {-s_{51}\over -s_{34}}\R  \over s_{34}^2} 
                                                 \label{FiveAmplsB}\\
      &&\null +{2\over 3} {{\spa2.3}^2 {\spa4.1}^3 {\spb2.4}^3
          \over {\spa4.5} {\spa5.1} {\spa2.4}}
          {\Ll_2\L {-s_{23}\over -s_{51}}\R  \over s_{51}^3}
      -{2\over 3} {{\spa2.1}^2 {\spa5.3}^3 {\spb2.5}^3
          \over {\spa5.4} {\spa4.3} {\spa2.5}}
          {\Ll_2\L {-s_{12}\over -s_{34}}\R  \over s_{34}^3} \cr
&&\null + {\Ll_2\L {-s_{34}\over -s_{51}}\R\over s_{51}^3}\,
     \biggl( {1\over3} { \spa1.3\spb2.4\spb2.5
     (\spa1.5 \spb5.2 \spa2.3-\spa3.4 \spb4.2 \spa2.1) \over \spa4.5}
  \cr
   & &\null +{2\over 3} {{\spa1.2}^2{\spa3.4}^2\spa4.1{\spb2.4}^3
            \over \spa4.5\spa5.1\spa2.4}
    -{2\over 3} {{\spa3.2}^2{\spa1.5}^2\spa5.3{\spb2.5}^3
            \over \spa5.4\spa4.3\spa2.5}  \biggr) \cr
  & &\null +{1\over 6} {{\spa1.3}^3
        \L \spa1.5\spb5.2\spa2.3 - \spa3.4\spb4.2\spa2.1\R
          \over \spa1.2\spa2.3\spa3.4\spa4.5\spa5.1}\,
            {\Ll_0\L {-s_{34}\over -s_{51}}\R\over s_{51}}
   +{1\over 3} {{\spb2.4}^2 {\spb2.5}^2
          \over {\spb1.2} {\spb2.3} {\spb3.4} {\spa4.5} {\spb5.1}}
            \cr
   &&\null -{1\over 3} {{\spa1.2} {\spa4.1}^2 {\spb2.4}^3
          \over {\spa4.5} {\spa5.1} {\spa2.4} {\spb2.3} {\spb3.4} s_{51}}
   +{1\over 3} {{\spa3.2} {\spa5.3}^2 {\spb2.5}^3
          \over {\spa5.4} {\spa4.3} {\spa2.5} {\spb2.1} {\spb1.5} s_{34}}
   +{1\over 6}
    {{\spa1.3}^2 \spb2.4\spb2.5 \over s_{34} \spa4.5 s_{51}}
           \ . \nn
\end{eqnarray}
In contrast to ref.~\cite{FiveYM}, in \eqns{FiveAmplsA}{FiveAmplsB}
 we use unrenormalized amplitudes;
this distinction actually has no effect on the corresponding gravity amplitudes 
since the difference drops out in \eqn{NeqFourSixMatter}.
The functions appearing in the above expressions are
\begin{eqnarray}
\Ll_0(r)&=& {\ln(r)\over 1-r}\,,\hskip 10mm
\Ll_1(r) = {\ln(r)+1-r\over (1-r)^2}\,,\hskip 10mm
\Ll_2(r) = {\ln(r)-(r-1/r)/2\over (1-r)^3}\,,\cr
\Ls_1(r_1,r_2) &=& \frac{1}{ (1-r_1-r_2)^2}
\Big{[} \Li_2(1-r_1) + \Li_2(1-r_2) + \ln r_1\,\ln r_2 - {\pi^2\over6} \\
   && \hskip 3 cm \null +(1-r_1-r_2)(\Ll_0(r_1) +\Ll_0(r_2))\Big{]} \,. \nn
\end{eqnarray}
As discussed in \sect{FivePtsGravSubsection}, these gauge-theory 
amplitudes serve as building blocks for the corresponding ${\cal N} \ge 4$ 
supergravity amplitudes.

\def\Fone{{\cal F}_1}
\def\Ftwo{{\cal F}_2}
\def\F#1#2{\,{{\vphantom{F}}_{#1}F_{#2}}}

\section{Integrals}
\label{IntegralsAppendix}

In this appendix we collect the integrals used in our expressions
from various sources and adjust normalization to match our
conventions.  The $m$-point scalar integrals in $D$ dimensions are
defined as:
\be \label{mPtsIntegral}
I_m= \int \frac{d^D p}{ (2 \pi)^D} \frac{1} {p^2 (p-K_1)^2
(p-K_1-K_2)^2 \ldots (p - K_1 - K_2 - \ldots - K_{m-1})^2}\,,
\ee
where the $K_i$'s are the external momenta which can be on- or off-shell.


The $D=4-2\epsilon$ bubble with momentum $K$ is
\be
I_2(K^2) = \frac{i c_{\Gamma}}{\epsilon(1-2\epsilon)}(-K^2)^{-\epsilon}\,,
\ee
where
\be
c_{\Gamma} = \frac{1}{(4\pi)^{2-\epsilon} }\frac{\Gamma(1+\epsilon) \Gamma^2(1-\epsilon)}{\Gamma(1-2\epsilon)}\,.
\label{cgamma}
\ee
The $D = 4 -2\epsilon$ one-mass triangle is 
\be
I_3(K_1^2) = \frac{-i c_{\Gamma}}{\epsilon^2}(-K_1^2)^{-1-\epsilon}\,,
\ee
where $K_1$ is the massive leg momentum and the two-mass triangle is
\be
I_3(K_1^2, K_2^2) = \frac{-ic_{\Gamma}}{\epsilon^2} \, 
\frac{ (-K_1^2)^{-\epsilon} - (-K_2^2)^{-\epsilon}}{ (-K_1^2) - (-K_2^2)} \,,
\ee
where $K_1$ and $K_1$ are the two massive leg momenta.

For amplitudes with four massless external particles we have the
zero-mass box $I_4^{1234} \equiv I_4(s,t)$ where $s = (k_1 + k_2)^2$,
$t = (k_2 + k_3)^2$ and the $k_i$ are massless momenta. An all-order
in $\epsilon$ expansion in terms of hypergeometric functions
is~\cite{DimRegPentagons}:
\begin{equation}
I_{4} (s,t) =  \ {2 i c_{\Gamma} \over \e^2 s t} \LB t^{-\epsilon}
   \F21\L -\epsilon,-\epsilon;1-\epsilon; 1+{t \over s}\R 
     \  + \ s^{-\epsilon}
   \F21\L -\epsilon,-\epsilon;1-\epsilon; 1+{s\over t}\R \RB \,,
 \label{ZeroMassBoxFull}
 \end{equation}
which through order $\epsilon^0$ is
 \begin{equation}
I_{4} (s,t) = \frac{ ic_{\Gamma}}{st}
\left[ \frac{2}{\e^2} \Big((-s)^{-\e}+(-t)^{-\e} \Big) -
\ln^2 \left(\frac{-s}{-t} \right)-\pi^2 \right] + \cal{O}(\epsilon) \,.
\label{ZeroMassBoxFullExpand}
\end{equation}
Similarly, the one-mass box through $\epsilon^0$ is~\cite{DimRegPentagons},
\begin{eqnarray}
 I_4^{(12)345} &=& -\frac{2ic_{\Gamma}}{s_{34}s_{45}} \Bigg \{ -\frac{1}{\epsilon^2}
\Bigl[ (-s_{34})^{-\e} + (-s_{45})^{-\e} - (-s_{12}^2)^{-\e} \Bigr]\nn \\
& &\null 
  + \Li_2\left(1-{s_{12}\over s_{34}}\right)
  +\Li_2\left(1-{s_{12}\over s_{45}}\right)
   + \frac{1}{2}\ln^2\left({s_{34}\over s_{45}}\right) + \frac{\pi^2}{6} \Bigg \}
+ \Ord(\e) \,, \hskip .5 cm 
\label{OneMassBoxIntegral}
\end{eqnarray}
where legs $1$ and $2$ are at the massive corner. An all orders in
$\epsilon$ form in terms of hypergeometric functions may be found in
ref.~\cite{DimRegPentagons}.  The integral $I_{4,\rm trunc}^{(12)345}$
is given by dropping the term multiplied by $1/\eps^2$,
\begin{equation}
I_{4, \rm trunc}^{(12)345}
= -\frac{2ic_{\Gamma}}{s_{34}s_{45}} \Bigg \{ 
 \Li_2\left(1-{s_{12}\over s_{34}}\right)
  +\Li_2\left(1-{s_{12}\over s_{45}}\right)
   + \frac{1}{2}\ln^2\left({s_{34}\over s_{45}}\right) + \frac{\pi^2}{6}\Bigg \}
+ \Ord(\e) \,.
\label{OneMassBoxIntegralTrunc}
\end{equation}
Finally, we use the pentagon integral whose expansion to order $\epsilon^0$ is \cite{DimRegPentagons}
\begin{eqnarray}
I_5^{12345}&=& \sum_{Z_5}{ -ic_{\Gamma} (-s_{51})^\eps (-s_{12})^\eps
   \over (-s_{23})^{1+\eps} (-s_{34} )^{1+\eps} (-s_{45})^{1+\eps} }
  \LB {1\over \eps^2} + 2\Li_2\Bigl(1- {s_{23} \over s_{51} }\Bigr)
  + 2\Li_2\Bigl( 1- {s_{45}\over s_{12}} \Bigr)
  - {\pi^2 \over 6} \RB  
 \nn \\
&& \null \hskip 5 cm 
+ \Ord(\eps)\,,
\end{eqnarray}
where the sum is over the five cyclic permutations of external legs.


We also need integrals in higher dimensions.  The triangle and bubble
integrals are obtained by direct integration and the box integrals by
dimension-shifting relations~\cite{DimRegPentagons}.  Explicitly, the
$D=6-2\epsilon$ bubble is
\be
I_2^{D=6-2\epsilon}(K^2) = \frac{-i c_{\Gamma}}{2\epsilon(1-2\epsilon)(3-2\epsilon)}(-K^2)^{1-\epsilon}\,,
\ee
whereas the $D=6-2\epsilon$ one-mass triangle is
\be
I_3^{D=6-2\epsilon}(K_1^2) = \frac{-i c_{\Gamma}}{2\epsilon(1-\epsilon)(1-2\epsilon)}(-K_1^2)^{-\epsilon}\,.
\ee
The zero-mass $D=6-2\epsilon$ box can be expressed as a linear
combination of the four-dimensional one-mass boxes and one-mass
triangles:
\be \label{6DboxFull}
I_4^{D=6 - 2\epsilon} (s, t) = \frac{1}{s+t} \, \Biggl(\frac{st}{2} I_4 (s,t) 
 - i \frac{c_\Gamma}{\e^2} \Big((-s)^{-\e}+(-t)^{-\e} \Big) \Biggr)\,.
\ee 
Note that it is finite and equal to
\begin{equation} \label{6DboxEps}
I_4^{D=6 - 2\epsilon} (s, t) =  -i \frac{ c_{\Gamma}}{ 2 (s+t)}
\left[\text{ln}^2\left(\frac{-s}{-t}\right) + \pi^2\right] + \Ord(\eps)\,.
\end{equation}

We also make use of the  integral combination from ref.~\cite{BernMorgan},
\begin{equation}
J_m = -\epsilon I_m^{D=6-2\epsilon}\,, \hskip 2 cm 
K_m = -\epsilon(1-\epsilon) I_m^{D=8-2\epsilon}\,.
\end{equation}
Through order $\epsilon^0$, these become
\be
J_4 = 0 + \Ord(\epsilon)\,, \hskip 5mm K_4 = -\frac{i}{6 (4 \pi)^2} 
   + \Ord(\epsilon)\,, \hskip 5mm J_3 = \frac{i}{2(4\pi)^2} +
 \Ord(\epsilon)\,.
\ee


\end{document}